# Optimisation of out-vessel magnetic diagnostics for plasma boundary reconstruction in tokamaks

J.A.Romero[1], J.Svensson[2]

[1]Laboratorio nacional de Fusión, Ciemat, Madrid,Spain

2 IPP Greisfwald, Germany

*E-mail contact of main author: jesus.romero@ciemat.es*

**Abstract**.

To improve the low frequency spectrum of magnetic field measurements of future tokamak reactors such as ITER, several steady state magnetic sensor technologies have been considered. For all the studied technologies it is always advantageous to place the sensors outside the vacuum vessel and as far away from the reactor core to minimize radiation damage and temperature effects, but not so far as to compromise the accuracy of the equilibrium reconstruction.

We have studied to what extent increasing the distance between out-vessel sensors and plasma can be compensated for sensor accuracy and/or density before the limit imposed by the degeneracy of the problem is reached. The study is particularized for the Swiss TCV tokamak, due to the quality of its magnetic data and its ability to operate with a wide range of plasma shapes and divertor configurations. We have scanned the plasma boundary reconstruction error as function of out-vessel sensor density, accuracy and distance to the plasma. The study is performed for both the transient and steady state phases of the tokamak discharge.

We find that, in general, there is a broad region in the parameter space where sensor accuracy, density and proximity to the plasma can be traded for one another to obtain a desired level of accuracy in the reconstructed boundary, up to some limit.

Extrapolation of the results to a tokamak reactor suggests that a hybrid configuration with sensors inside and outside the vacuum vessel could be used to obtain a good boundary reconstruction during both the transient and the flat-top of the discharges, if out-vessel magnetic sensors of sufficient density and accuracy can be placed sufficiently far outside the vessel to minimize radiation damage.

## 1. Introduction

Nuclear fusion reactors based on the tokamak concept require a set of magnetic probes to reconstruct the plasma equilibrium that can measure magnetic field and at the same time are able to withstand harsh conditions of radiation with sufficient temperature stability and accuracy. These sensors are primarily used to infer the plasma current, boundary shape and position and to provide this information to magnetic control, whose function is to keep the hot plasma away from the vacuum vessel walls, using the currents in the poloidal field PF coils as the main actuators [1]-[14] . So far, tokamaks have used conventional Mirnov coils (also called pick-up coils) to measure magnetic field time variation dB/dt. The magnetic field is then obtained by time integration of the signal using the proper calibration. The approach has been adequate for tokamaks used in plasma physics experiments, with low levels of radiation. However, as reactor conditions are approached, some difficulties arise. Measurement errors arising from radiation induced thermoelectric effects (RITE) in the signal transmission cables (which take the form of spurious spikes) lead after time integration to offsets in the magnetic measurements that are difficult to correct, so the low frequency spectrum of the magnetic field is not well determined in long pulse steady state scenarios [3]-[5].

To correct/replace these measurements and/or improve the low frequency spectrum of magnetic field measurements, steady state magnetic out-vessel sensors not requiring time integration are being sought after for the projected International Thermonuclear Fusion Reactor ITER [6].

Several steady state sensing technologies have been studied in the context of nuclear fusion. One of the first technologies being considered is Hall effect sensors [7] -[10]. Because these sensors are based on semiconductor technology, they are prone to electron/hole mobility changes due to radiation induced lattice damage, which alters the transport properties of the sensor (e.g. Hall coefficient) and requires periodic calibrations. They also need a controlled temperature environment or compensation circuits to achieve sufficient temperature stability of the measurements.

To avoid the radiation damage and change on sensor properties, Perovskite based Colossal Magneto Resistance (CMR) sensors have also been studied [11]. The advantage of CMR sensors is that there is indication of their transport properties (e.g. electrical conductivity as function of magnetic field) remaining largely unchanged despite large damages to the perovskite lattice. The disadvantage is that CMR sensors work close to the Curie temperature of the material, so they are also highly sensitive to temperature changes, requiring a controlled temperature environment.

Yet another type of out-vessel sensors studied in the context of nuclear fusion research is based on Micro Electro Mechanical systems (MEMS) [12]. MEMS magnetic sensor can be configured to be almost immune to the change in elastic properties of its elements being induced by radiation damage. However they require the electronics to be very close to the sensing elements. If semiconductor based, the electronics is in turn sensitive to radiation and temperature conditions.

For all the above sensing technologies, it is always advantageous to place the sensors as far away as possible from the reactor core to minimize radiation damage and temperature effects, but not so far as to compromise the accuracy of the equilibrium reconstruction. We wonder to what extent increasing the distance between sensors and plasma can be compensated by sensor accuracy before the limit imposed by the degeneracy of the problem is reached. Regardless of the out-vessel sensor technology used, an analysis of how far away these sensors can be placed as function of their required accuracy is worthy.

This work investigates if out-vessel sensors placed away from the plasma volume outside the main vessel can be used to gain additional information on the plasma boundary in nuclear fusion tokamak devices. We are interested, in particular, if sensor accuracy and distance to the plasma can be traded for one another, meaning at what range of positions these sensors could be placed as function of their measurement uncertainty in order to get the desired plasma boundary reconstruction accuracy.

We have based the study on the *Tokamak à Configuration Variable* (TCV) experiment [13],[14] in Lausanne, Switzerland, on account of its rich variety of plasma shapes, divertor configurations and magnetic diagnostics. For this device there is an extensive magnetic database that can be used for the analysis [15], [16].

The study involves a full model of the TCV magnetic sources (section 2), sensors (section 3) and a linear model that relate both (section 4). We have used Bayesian inference [17] reconstruction methods (current tomography) on account of its ability to provide an uncertainty measure on the reconstructed boundary (section 5). Current tomography (CT ) is a well established method successfully used in JET [18] and MAST [19] tokamaks. The reconstruction algorithm involves only linear operations, so it is also suitable for real time computation. It gives full uncertainties on the reconstructed boundary given the magnetic measurement uncertainty, the expected range of current induced in the vacuum vessel, and the length scales for correlations among the different elements of the plasma current distribution (section 6). A specific application for the TCV machine has been deployed using the MINERVA framework [20], and used for the analyses in this paper. It uses Bayesian probability theory both for the inference of maximum probable boundary given the measurements, and for inferring the uncertainties in the reconstructed boundary. The length scales for plasma current distribution and vessel current intensity (section 6) are inferred using a Bayesian evidence optimization approach [25], which is discussed in section 7. The equilibrium database used is described in section 8. The analysis is restricted to the quality of the boundary reconstruction, defined in section 9.

The following sections analyse a subset of the equilibria in the database to uncover the mayor trends in steady state (section 10) and transient (section 11) equilibria.

A full parameter scan of the boundary reconstruction error as function of sensor density, accuracy and distance to the plasma is presented in section 12, using models valid for both transient and steady state situations. This is followed by a discussion and the main conclusions of this work.

## 2. TCV magnetic field model

The TCV magnetic field model used in the analysis comprises 20 active systems (e.g. coil systems connected to their power supplies) and two passive systems (Fig 1). The active systems comprise the poloidal and toroidal field coils. The poloidal field system comprises the transformer primary coil (A), plasma shaping coils (B-F) and internal coils for fast position control (G) [14] , [15]. These coils are modelled using current beams of rectangular cross-section carrying a uniform current proportional to the number of coil turns, as summarized in table 2, and grouped in active coil systems according to table 3 (see appendix). The toroidal field is generated by a discrete set of 16 equally spaced toroidal field coils. The feeding lines for the toroidal field coils run toroidally and thus produce a small poloidal field, which is treated as an additional poloidal field coil circuit labelled T_001. This comprises three poloidal coils labelled T_001, T_002 and T_003. The current follows 26/68 of a whole turn in the T_001 coil and the remaining 42/68 in the T_002 coil. The return current is carried by the complete turn coil T_003. This is reflected in last row of table 3 of the appendix.

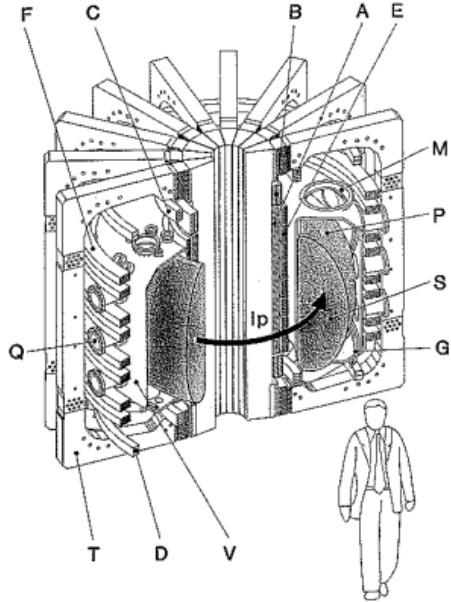

Fig. 1. TCV machine. A: Ohmic transformer coil, B-F: poloidal field coils for plasma shaping, G: internal poloidal field coil for fast position control, M: large port for maintenance access, Q: diagnostic ports, P: highly elongated plasma with current Ip , S: hot central region of the plasma, T: toroidal field coils, V: vacuum vessel

The passive system comprises the vacuum vessel and the plasma. The currents induced in the vessel due to changes in poloidal flux are modelled by a set of 256 thin (0.005x0.005 m) solid beams along the vacuum vessel perimeter. The plasma current is modelled by a grid of 30x33 current beams of rectangular cross-section covering an area inside the first wall. The final arrangement including passive and active system can be appreciated in fig 2.

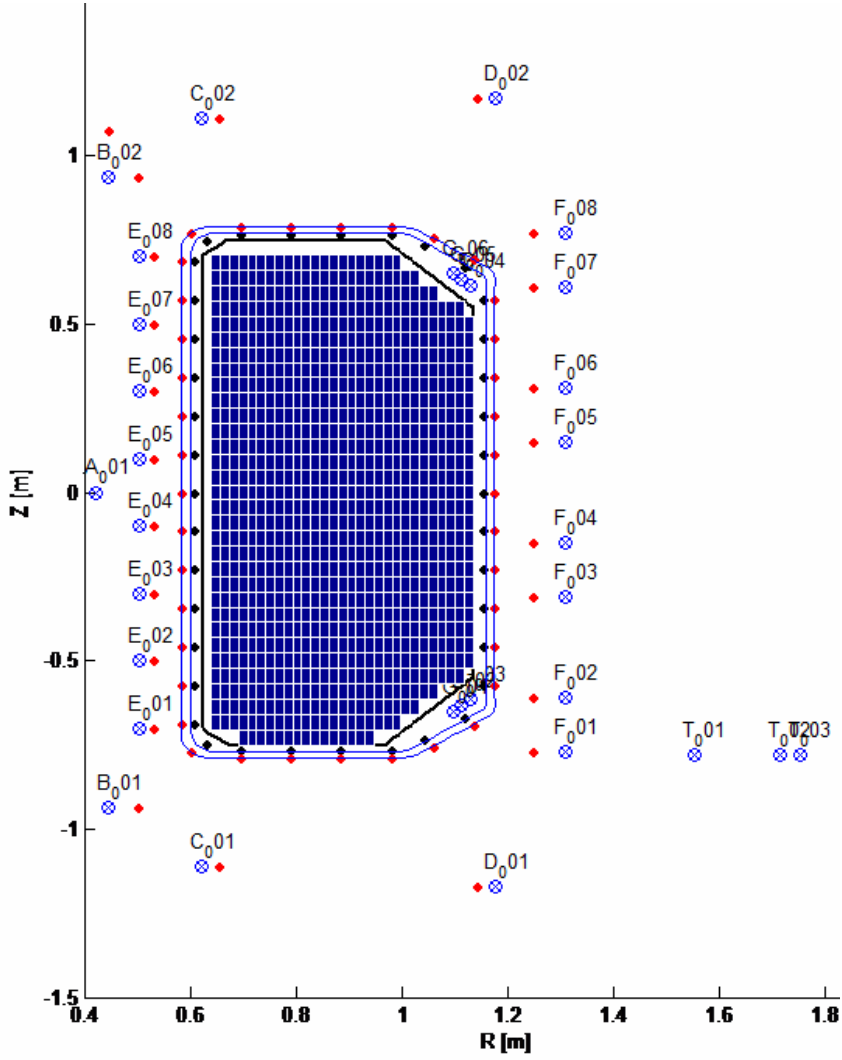

Fig. 2. Plasma toroidal current beam grid (solid blue) together with centre points for PF coils (blue circles with crosses), flux loops (red circles), and Mirnov coils (black circles). 256 vessel current beams (not shown) are uniformly distributed inside the vacuum vessel walls (concentric blue lines).

Using the Biot-Savart law for magnetic field and vector potential, the contribution to the signal at each measurement position, from all the solid beams constituting our magnetic model, is then calculated, using both active and passive systems.

In the following, a cylindrical coordinate system $(r, \phi, z)$ is used, and the plasma is assumed to be axis-symmetric about the z-axis.

Using the vector potential with Coulomb gauge

$$\bar{A} = \begin{pmatrix} A_r & A_\phi & A_z \end{pmatrix} = \begin{pmatrix} 0 & \dfrac{\psi}{2\pi r} & 0 \end{pmatrix} \tag{1}$$

where $\psi$ is the flux through an arbitrary circle of radius r centred at the torus symmetry axis, the magnetic field can be obtained from a vector potential, as

$$\bar{B} = \nabla \times \bar{A} \tag{2}$$

This renders the usual expressions for magnetic field components in a tokamak

$$B_r = -\frac{\partial A_\phi}{\partial z} = -\frac{1}{2\pi r}\frac{\partial \psi}{\partial z}$$
$$B_z = \frac{1}{r}\frac{\partial \left(r A_\phi\right)}{\partial r} = \frac{1}{2\pi r}\frac{\partial \psi}{\partial r}$$

(3)

The toroidal component of the field is calculated as function of the total current in the toroidal field coils $I_{tor}$ by an axis-symmetric field with a 1/R dependency so no 3D effects are considered.

$$B_\phi = \frac{\mu_0 I_{tor}}{2\pi r}$$

(4)

The vector potential relates to the toroidal current density $j_\phi$, [A/m2] in all the active and passive systems through Biot-Savart's law

$$A_\phi = \frac{\mu_0}{4\pi}\iiint \frac{j_\phi(\overrightarrow{r'})}{\left|\overrightarrow{r}-\overrightarrow{r'}\right|}dV'$$

(5)

where r is a field point and r' a source point.

It is clear that the poloidal flux, which determines the plasma boundary, is a second integral of the toroidal current distribution, and so is relatively insensitive to small local changes in the current distribution and a moderately high resolution grid for the plasma current distribution can therefore be used for the feasibility study. Small adjustments of the plasma currents in the individual beams will smoothly change the position of the plasma boundary since the poloidal flux in itself is not discretized.

## 3. TCV magnetic field sensors

The TCV magnetic sensor set is described in detail in [15], and is only partially summarized here. It comprises a set of 38 Mirnov coils placed inside the vessel and giving the local magnetic field in a direction tangential to the vacuum vessel, with positions and directions given in table 4 of the appendix. The full poloidal flux is measured by a number of flux loops that run toroidally and are attached to the outer PF coils insulation (coil loops) and by a number of both flux loops and saddle loops that extend toroidally and are positioned just outside the vacuum vessel (vessel loops). Not all of the vessel loops run on a constant radius around the vacuum vessel, nor all of them complete a full toroidal turn, since some of them must bypass the vacuum vessel ports.

From all the vessel loops, a set of ideal flux loops 1-38 surrounding the vacuum vessel have been reconstructed from measurements in a preprocessing procedure, and stored as the flux at 38 locations in the TCV database. 1-38 are then indexes to the virtual flux loops which have been reconstructed, and do not always correspond with actual installation.

These ideal flux loops are the ones used for this study together with the flux loops 39-61 placed near the PF coils, named coil flux loops. The positioning of all these flux loops is summarized in tables 5,6 of the appendix. With this system, accuracy of 0.5 mWb in the poloidal flux and 1 mT in the magnetic field with a position error of a few mm have been achieved.

The PF coils currents are measured with zero flux transformers which have a relative accuracy of 0.25% of the nominal coil current, which corresponds to about 100 A for the OH currents and 20 A for the E and F currents.

## 4.    Forward model

The free parameters for the problem being investigated are the set of current densities of the plasma $j_\phi^i$ and vessel $j_v^j$ beams arranged in a vector $\overline{I}$

$$\overline{I} = \{\{j_\phi^i\}, \{I_v^j\}\} \tag{6}$$

Similarly, a data vector $\overline{D}$ will include combinations of observations from Mirnov coils $m_i$, flux loops $f_j$ and out-vessel sensors $s_k$, depending on the study being made.

$$\overline{D} = \{\{m_i\}, \{f_j\}, \{s_k\}\} \tag{7}$$

The forward model uses the Biot Savart operator $f$ to map the different current densities in $\overline{I}$ to predicted measurements $f(\overline{I})$ of the Mirnov coils, flux loops and out-vessel sensors. From magneto-static theory it follows that these quantities are linearly related and can therefore be described by the matrix expression

$$f(\overline{I}) = \overline{\overline{M}}\,\overline{I} + \overline{C} \tag{8}$$

where $\overline{\overline{M}}$ is an $N_D \times N_I$ matrix, being $N_D$ is the number of measurements and $N_I$ the number of free parameters. $\overline{C}$ is the background contribution from the PF coil system to each measurement, with elements $C_i$ corresponding to the contribution to measurement i from the PF coils, assuming zero plasma and vessel currents. The coefficients in $\overline{\overline{M}}$ contain the contributions to the signal at the different measurement positions from the different solid beams constituting our magnetic model. $M_{ij}$ is the contribution to measurement i from current beam j for a unit current or unit current density.

## 5.    The inversion method

 The inversion method to obtain the parameters $\overline{I}$ from the data vector $\overline{D}$ is based on the Bayesian inference methods developed in [18],[25]. The main reason to choose this inversion method in particular is that compared with other available methods it is the only method that provides an uncertainty measure on the reconstruction.

In principle, there is a large family of internal plasma current distributions compatible with a given magnetic data. The solution space for $\overline{I}$ that is compatible with the observations $\overline{D}$ can be arranged into a probability distribution function $p(\overline{I}\,|\,\overline{D})$ called the posterior distribution. The posterior distribution $p(\bar{I}\,|\,\bar{D})$ is the end result of inference on the free parameters $\bar{I}$, given a specific data vector $\bar{D}$. It fully describes all uncertainties associated with measurement errors on $\bar{D}$, and inherent ambiguities in the solution space $\bar{I}$, as demonstrated in [18].

According to Bayes formula [17]

$$p(\bar{I}\,|\,\bar{D}) = \frac{p(\bar{D}\,|\,\bar{I})\,p(\bar{I})}{p(\bar{D})} \tag{9}$$

where $p(\bar{D}\,|\,\bar{I})$ is the likehood probability distribution, $p(\bar{I})$ is the prior distribution (a measure of our a priori knowledge about the unknown currents, to be determined as explained below) and the term in the denominator $p(\bar{D})$ is called the evidence, and normalizes the volume of the posterior distribution to 1.

$$p(\bar{D}) = \int_{...} p(\bar{D}\,|\,\bar{I})\,p(\bar{I})\,d\bar{I} \tag{10}$$

A main result obtained in [18] is that the mean values for the current in the plasma and vacuum vessel is given as an explicit function of the data vector $\bar{D}$, data uncertainties expressed $\bar{\bar{\Sigma}}_D$, and prior covariance $\bar{\bar{\Sigma}}_I$.

$$\bar{m} = (\bar{\bar{M}}^T \bar{\bar{\Sigma}}_D^{-1} \bar{\bar{M}} + \bar{\bar{\Sigma}}_I^{-1})^{-1} \bar{\bar{M}}^T \bar{\bar{\Sigma}}_D^{-1} (\bar{D} - \bar{C}) \tag{11}$$

and the uncertainty of the solution is given by the posterior covariance

$$\bar{\bar{\Sigma}} = (\bar{\bar{M}}^T \bar{\bar{\Sigma}}_D^{-1} \bar{\bar{M}} + \bar{\bar{\Sigma}}_I^{-1})^{-1} \tag{12}$$

The data covariance matrix $\bar{\bar{\Sigma}}_D$ is used to parametrize the misfit between the model $f(\bar{I})$ and the data $\bar{D}$, which is described by the likehood distribution $p(\bar{D}\,|\,\bar{I})$. The likelihood distribution is taken as a normal distribution with a given standard deviation for each observation:

$$p(\bar{D}\,|\,\bar{I}) = \frac{1}{(2\pi)^{N_D/2}\left|\bar{\bar{\Sigma}}_D\right|^{1/2}} \exp(-\frac{1}{2}(f(\bar{I}) - \bar{D})^T \bar{\bar{\Sigma}}_D^{-1}(f(\bar{I}) - \bar{D})) \tag{13}$$

The covariance matrix of the data $\bar{\bar{\Sigma}}_D$ is diagonal with $\Sigma_D(i,i) = \sigma_i^2$ where $\sigma_i$ is the uncertainty for measurement i.

In case of TCV, for instance, $\sigma_i$ is 0.1mT for Mirnov coils and 0.5mWb for flux loops.

Likewise the prior covariance matrix $\bar{\bar{\Sigma}}_I$ is used to parameterize the prior distribution, and serves as a regularising constraint for the tomographic inversion problem. The prior distribution is taken as a multivariate normal distribution over the currents and current densities in the form

$$p(\overline{I}) = \frac{1}{(2\pi)^{N_I/2} \left| \overline{\overline{\Sigma}}_I \right|^{1/2}} \exp(-\frac{1}{2}(\overline{I} - \overline{m}_I)^T \overline{\overline{\Sigma}}_I^{-1} (\overline{I} - \overline{m}_I)) \tag{14}$$

where $\overline{m}_I$ is the prior mean (which will here be taken as zero), and $\overline{\overline{\Sigma}}_I$ is the prior covariance.

It is convenient but by no means necessary to consider a zero mean on the prior. Setting zero mean allows to treat equally positive and negative beam current values, so we make no assumption about the sign of the current. Note that this is not necessarily a drastic limitation, since the mean of the posterior (Eq. (11)) is not confined to be zero.

The diagonal elements of $\overline{\overline{\Sigma}}_I$ contain the variances for each variable, while the off-diagonal elements of $\overline{\overline{\Sigma}}_I$ contain the covariances between variables. This high dimensional covariance matrix provides a very flexible way of expressing prior regularizing constraints. The elements of the prior covariance matrix $\overline{\overline{\Sigma}}_I$ for both the vessel currents and the plasma current beams are determined directly from the experimental data, before the actual tomography analysis, following the method described in following sections 6,7.

By putting the expression for the likelihood (13) and the prior (14), into Bayes formula (9) and rearranging terms, the posterior distribution can be written as a multivariate normal distribution $p(\overline{I} \mid \overline{D})$. The most likely single solution (the maximum posterior estimate) of this distribution is given by (11), with its uncertainty given by (12).

Once the plasma current distribution and vessel currents are determined, poloidal flux is calculated using the model of section 2, and flux surfaces determined using contouring algorithms. The actual inversion is a two step process: in the first step the last closed flux surface is found using all the beams in (6), then the inversion is done again keeping only the plasma beams inside the contour found in the first step. This method implicitly enforces the prior knowledge that there is no significant current between the last closed flux surface and the vacuum vessel, and improves the boundary reconstruction by reducing the space of possible solutions. The most probable boundary will correspond to the maximum posterior estimate for the current distribution (11). Once this is determined, flux and flux contours are calculated a number of times by exploring a wide range of solutions near (11) that are compatible with the measurements. This is done by taking samples of the high dimensional posterior distribution (9), reconstructing the boundary for each sample. Calculating misfit measures such as standard deviation from the maximum probable boundary, uncertainties of the boundary can be estimated (see Fig. 3 for a block diagram). This procedure takes into account the effect of the posterior uncertainties of the unknown vessel currents, which will contribute to the uncertainty of the boundary.

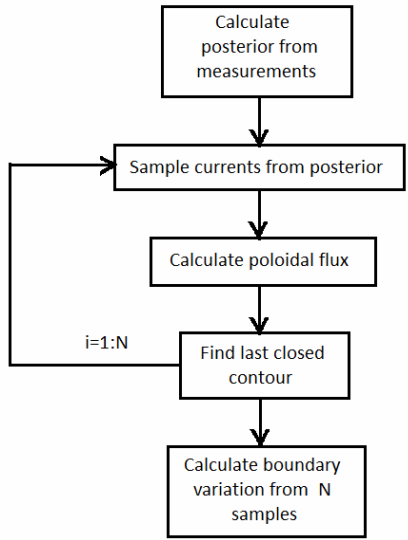

Fig. 3. Block diagram of procedure to calculate uncertainty of boundary reconstruction.

As we explained before, we obtain a large family of current profiles which are compatible with the magnetic measurements in the form of a probability distribution (the posterior distribution). The most likely single solution (the maximum posterior estimate) of this distribution is an average value given by (11), but this average profile doesn't need to correspond with a Grad-Shafranov consistent equilibrium profile. The CT method gives a posterior probability distribution over all possible toroidal current profiles compatible with the magnetic data and the prior assumptions. Reducing the uncertainty (12) of the current profile solution always requires additional internal diagnostics such as MSE, and/or equilibrium constraints, forcing the posterior distribution to be compatible not only with magnetic diagnostics but also with these internal current profile observations. Since the focus of the paper is the boundary reconstruction and its accuracy, we are not discussing in the paper the quality of the full equilibrium. Within the scope of this work, the use of current distributions can be viewed as a step stone used to reconstruct the plasma boundary and its uncertainty. There are other published works where the issue of full equilibrium reconstruction with CT using polarimetry [21] and equilibrium constraints [22], [23] is discussed.

## 6. Modelling the prior as a Gaussian Process

A Gaussian process GP is a collection of random variables, any finite number of which has a joint Gaussian distribution. A Gaussian process is completely specified by its mean function and co- variance function. The application of Gaussian processes to Bayesian analysis has been explored in several different contexts, such as regression, classification problems in the pattern recognition field and machine learning [24] .

The elements $\sigma_{ij}^2$ of the prior covariance corresponding to every pair of toroidal plasma current beams can be parameterized using the expected length scales for plasma current density and variation of the plasma current distribution in the R and Z directions, using the so called squared exponential GP [24] :

$$\sigma_{ij}^2\left(\left(R_i, Z_i\right),\left(R_j, Z_j\right)\right) = \sigma_f^2 \exp\left(-\frac{\left(R_i - R_j\right)^2}{2\sigma_R^2} - \frac{\left(Z_i - Z_j\right)^2}{2\sigma_Z^2}\right) \tag{15}$$

the parameters $\sigma_f$ , $\sigma_R$ and $\sigma_Z$ in the GP context are called hyper parameters . They define the length scales of the current density ($\sigma_f$), the length scale in the R direction ($\sigma_R$), and the length scale in the Z direction ($\sigma_Z$).   Note the impossibility to reduce both scale lengths to a single scale length in poloidal flux space, since the toroidal current density is not a flux function.

The regularization imposed by this prior can be understood from inspection of (15) : a small length scale in the R-direction (small $\sigma_R$) will give lower covariance $\sigma_{ij}$ between beams with different R coordinates in comparison to using a larger $\sigma_R$. In other words, if the current profiles are smooth, the corresponding scale lengths will assign a low probability for large differences between neighbouring current beams. This prior and corresponding method has previously been used to successfully reconstruct boundary and internal flux surfaces on the JET tokamak [25].

The vessel current prior distribution, which is part of (14), is a multivariate normal distribution with a single variance hyper parameter $\sigma_v^2$ :

$$p(\overline{I}_v) = N(0, \sigma_v^2) = \frac{1}{(2\pi)^{N_V/2}\sigma_V^{N_V}}\exp(-\frac{1}{2\sigma_V^2}\sum_{i=1}^{N_V}I_{Vi}^2) \tag{16}$$

where $I_{Vi}$ is vessel current i.   The full prior covariance matrix $\overline{\overline{\Sigma}}_I$ in (14) is formed from the prior covariance elements for the plasma beam densities in (15), and the diagonal covariance from the vessel currents prior (16).

## 7.  Determination of the prior distribution hyper parameters

If we assume the hyper parameters of section 6 ($\sigma_f, \sigma_R, \sigma_Z$) are fixed, Bayes formula (9) can be written conditioned on those hyper parameters:

$$p(\overline{I}\mid\overline{D},\sigma_f,\sigma_R,\sigma_Z) = \frac{p(\overline{D}\mid\overline{I},\sigma_f,\sigma_R,\sigma_Z)\,p(\overline{I}\mid\sigma_f,\sigma_R,\sigma_Z)}{p(\overline{D}\mid\sigma_f,\sigma_R,\sigma_Z)} \tag{17}$$

and we see that the evidence term in the denominator acquires the form of a likelihood term for the data $\overline{D}$, where the free parameters have been integrated out (10). In this form, the evidence term functions as a likelihood for the hyper parameters themselves, and by writing down Bayes formula for the hyper parameters using this likelihood

$$p(\sigma_f,\sigma_R,\sigma_Z\mid\overline{D}) = \frac{p(\overline{D}\mid\sigma_f,\sigma_R,\sigma_Z)\,p(\sigma_f,\sigma_R,\sigma_Z)}{p(\overline{D})} \tag{18}$$

we can estimate the values of the hyper parameters from given data $\overline{D}$. The expression (18) for the posterior of the hyper parameters can be calculated analytically (the integral in (10) can be carried out for fixed hyper parameters), since the forward model (8) is linear, the likelihood (13) is normal, and the prior (14) is also normal [18]. We will use a uniform prior on the hyper parameters, in which case (18) is proportional to the evidence (10) as a function of the hyper parameters.

By nonlinearly optimizing this evidence term respect to the hyper parameters (or in other words, finding the maximum posterior for the hyper parameters with uniform priors) we can determine the maximum likely hyper parameters for the study, providing us with an objective method of choosing the hyper parameters of the regularizer (15), as is shown in the following examples.

There are three hyper parameters $\sigma_f, \sigma_R, \sigma_Z$ for the plasma prior Gaussian process (15). $\sigma_f$ regulates the magnitude of the plasma current density, and can be set manually from knowledge about the approximate average magnitude of the TCV plasma current density at any given beam position (including all beams inside the vessel). For this study we have set $\sigma_f = 0.4 MA/m^2$. This parameter sets the rough scale (prior variance) of the current density itself, for an arbitrary beam, and can be set manually to an expected current density at an arbitrary beam position, for a given machine. For the parameters $\sigma_R, \sigma_Z$ which determine the length scale of variation of the plasma current distribution in the R and Z directions, a more careful study has to be done. These are estimated directly from the magnetic data, by optimizing the evidence $p(\overline{D} | \sigma_R, \sigma_Z)$ in the space of $\sigma_R, \sigma_Z$ using a Hooke & Jeeves [26] nonlinear optimizer. The magnetic data comes from pulse 39950, a plasma current modulation experiment chosen for its large and fast variation of plasma current [27] , which functions here as a calibration pulse for the hyper parameters.

In the chosen discharge, the current modulation raise time (30ms) is several times faster than the plasma current profile diffusion time (L/R = 200 ms). Because steady state conditions are not fully reached, it potentially includes lot of plasma current profile variability, which helps in finding the length scales for a large set of plasma current distributions. Likewise the discharge is expected to have a substantial amount of induced current in the vacuum vessel, which is required to find the vessel current hyperparameters.

The length scales are estimated at a number of different times through the pulse, and the average value of the hyper parameters will then be used for the rest of the study. This is similar to the "history prior" approach used in [18], but is less restrictive and calibration can be done on a smaller dataset. Fig 4 shows the variation of the optimised hyper parameters over the pulse. The average values found for the length scales are $\sigma_R = 0.16m$, $\sigma_Z = 0.18m$. This method then gives a measure of the length scales as assumed when TCV pulses are normally analysed, which is what we will be using for this study.

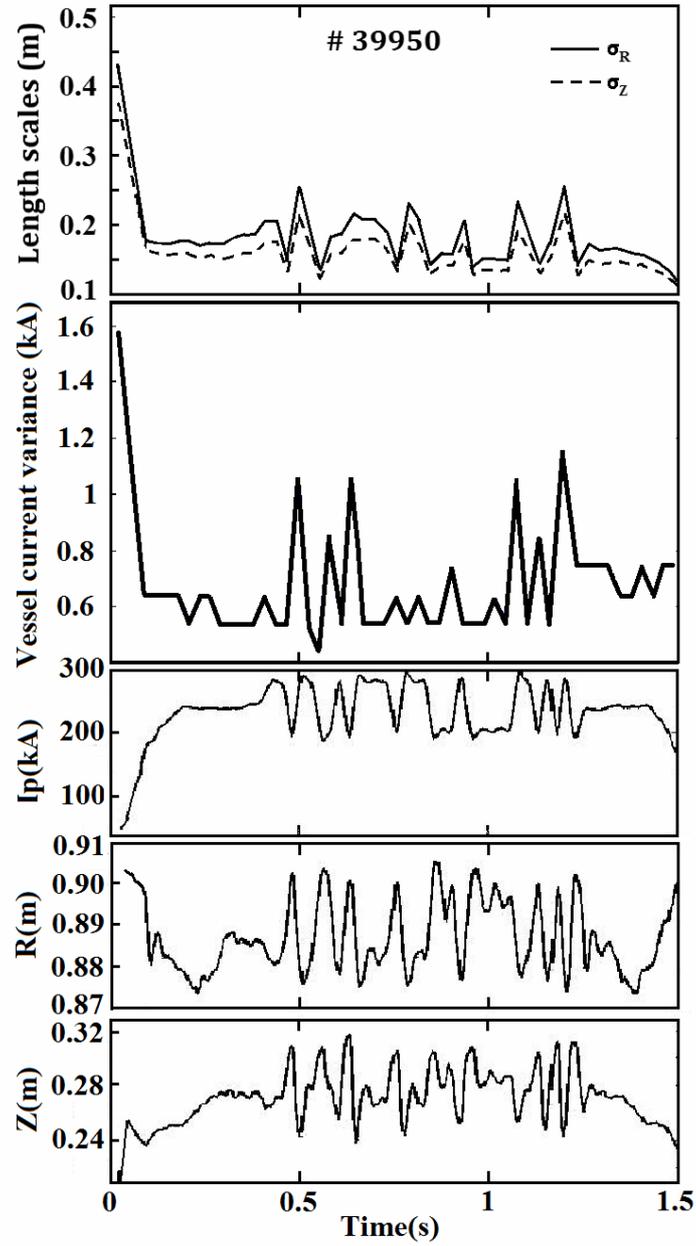

Fig. 4. From the top: Optimization of length scales of plasma current distribution in the R and Z directions, vessel current variance, plasma current and plasma position R,Z. The current distributions come from pulse 39950, a plasma current modulation experiment chosen for its large and fast variation of plasma current. Averages over the whole pulse are $\sigma_R = 0.16m$, $\sigma_Z = 0.18m$.

The model for the vessel currents has 256 current-carrying beams which would without regularization overfit the solution. The single hyper parameter $\sigma_v^2$ of (16) will be determined by calculating the evidence $p(\overline{D}|,\sigma_v)$ for $\sigma_v^2$ over pulse 39950 on the actual measurements $\overline{D}$ of the Mirnov coils and flux loops. Fig 4 shows the optimised vessel current variance over the calibration pulse by maximization of the evidence for different times over pulse 39950. The maximization of the evidence has here been done by scanning $\sigma_v$ over [0.01, 1.5] for each time slice, and picking the $\sigma_v$ corresponding to maximum evidence.

## 8.  Equilibria and simulated measurements database

The TCV contains extensive information from magnetic diagnostics corresponding with a large variation of plasma shapes and divertor configurations. For the purpose of this study, a subset of 383 equilibria taken from 13 different TCV pulses have been used. Table 7 of the appendix shows the pulses in the full database together with time limits within those pulses within which equilibria has been randomly selected. These not only explore different plasma shapes, but also plasma current modulation experiments [27] where a significant amount of current is expected to be transiently induced in the vacuum vessel. These are used to test the influence of the vessel currents in the reconstructions. A smaller subset drawn randomly from this database has been used for most of the study. The reason for using a smaller subset for most trend analysis is for reasons of efficiency and the possibility of direct inspection in the study of the actual plasma shapes used. When necessary, the larger database has then been used for verification.

Using this database, two different types of solutions are generated from the information in Mirnov coils and flux loops. In one class of solutions only plasma current density distribution is inferred. In the other class of solutions, both plasma and vessel currents are inferred. The maximum posterior estimates for both classes are used as reference cases. The corresponding last closed flux surface is likewise used as a reference boundary. From the inferred vessel currents, if used, and given toroidal current and PF circuit currents, the measurements that a particular magnetic sensor ( Mirnov, flux loop or out-vessel sensor) would have picked at any desired spatial location can then be simulated using our magnetic model. Various normally distributed noise levels corresponding with the assumed sensor uncertainty are then added to these simulated measurements. The result is a database of equilibria test cases containing the reference boundary along with the corresponding simulated magnetic measurements contaminated with different levels of noise. These are later used to infer again the plasma boundary. The comparison of the inferred boundary with the reference boundary gives a boundary reconstruction error.   For the out-vessel sensors, the database contains an accuracy range from $1\,\mu$T to 1mT and a distance range 5cm to 60 cm away from the Mirnov coil position, well outside the vacuum vessel.   The placing perimeters for the out-vessel sensors are shown in fig. 5 for a range from 5cm to 20cm.

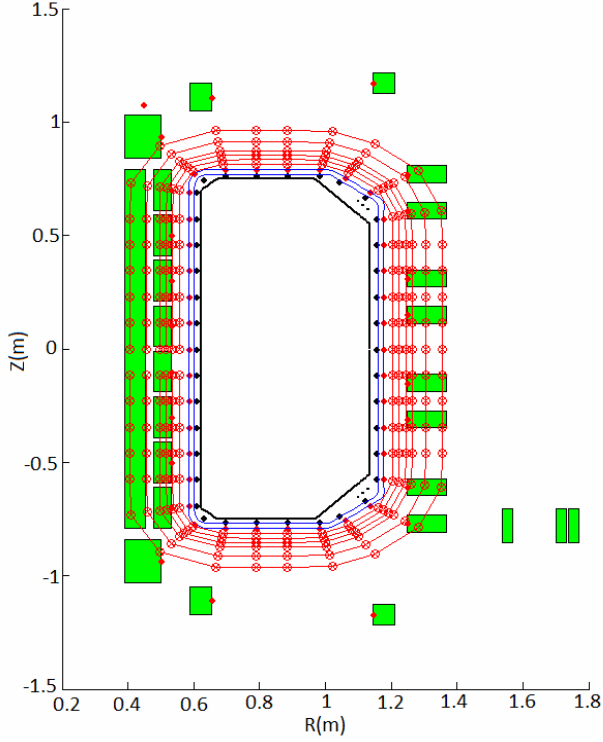

Fig. 5. Out-vessel sensor positioning (crossed circles in red) and placing perimeters (in red) for distances from Mirnov coils d=0.05, 0.07, 0.09, 0.11, 0.15, 0.2 m. Flux loops and Mirnov coil positions are marked with red and black solid dots respectively.

## 9. Evaluation of boundary reconstruction quality as function of magnetic sensors accuracy and positioning

All studies have been done on a simulated database as described in the previous section. The simulated measurements with added noise are used to infer the posterior distribution of plasma and vessel currents, depending on the test case. From here the poloidal flux is calculated and last closed contour is determined.

To calculate the misfit, 100 evenly distributed radial chords starting at the geometric centre of the reference boundary are defined, spaced by 3.6 degrees in the poloidal angle. The intersection of these radial cords with the boundary defines a set of 100 points along the boundary. The inferred boundary is compared with the true boundary at these positions, and a boundary reconstruction error is determined. The misfit measure for a single boundary reconstruction $j$ in the database is given by

$$e_j = \frac{1}{100} \sum_{i=1}^{100} \left| d_i^{tb} - d_i^{ib} \right| \tag{19}$$

where $d_i^{tb}$ is the distance between the centre of the boundary polygon and the reference boundary point along chord i, and $d_i^{ib}$ the distance between the same chord and the inferred boundary. The average misfit for a single boundary reconstruction $e_j$ is then the mean of the absolute value of the difference between these two distances over all chords, and is here usually expressed in millimetres.

The procedure is repeated for all the equilibria in the database and a single misfit measure is produced:

$$E_j = \frac{1}{N}\sum_{j=1}^{N} e_j \qquad\qquad\qquad (20)$$

where N is the number of boundary reconstructions in the database used.

For this study, when we talk about uncertainties, accuracies or errors of sensor measurements, we will be referring to the accuracy with which our model can predict the actual sensor value, and so could possibly be more accurately called "model prediction accuracy". From the point of view of Bayesian practice, this is what constitutes the error in relation to a measurement, and this will include assumptions, or tests, on how well our model can predict a given measurement. It is therefore a more full description of how much information a given measurement can provide on our (always imperfect) model, than a given intrinsic error of a sensor. Since the expression "sensor accuracy" is so engrained in standard instrumentation vocabulary this expression has been kept in some places, but it should be kept in mind that it is the predictive capacity of our model in explaining the measurement that is meant. How accurate the model is can only be checked once out-vessel sensors are actually installed in a real system. The model prediction accuracy could be checked by for example measuring the out-vessel sensor responses during dry runs, to check the PF model, sensor alignment and toroidal field compensation accuracy.

With the standard Mirnov and flux loop sensors, because of their much higher intrinsic measurement error ($10^{-3}$ Tesla in comparison to the Hall probes $10^{-4}$ Tesla), it is not possible to check the model prediction accuracy to a higher accuracy than that given by the sensor error. The advantage of using more accurate out-vessel sensors such as Hall probes is that the intrinsic measurement error is so low so that it will not be the limiting factor of the prediction accuracy.

## 10. Analysis examples with no current induced in the vacuum vessel

A magnetic sensor configuration is defined in terms of the type of sensor (bidirectional sensors or flux loops), the positioning (distance to the plasma, in or out-vessel), density (number of sensors) and sensor accuracy (or measurement uncertainty). We would expect that for a given, fixed magnetic sensor configuration (in terms of sensor density and positioning), the boundary reconstruction error would decrease as sensor accuracy is increased. We would similarly expect that for a given sensor configuration (in terms of sensor accuracy and density), the boundary reconstruction error would decrease as sensors are located closer to the plasma. We would finally expect that for a given sensor configuration (in terms of sensor accuracy and positioning), the boundary reconstruction error would decrease as sensor density is increased. To have an initial approximation to the problem, in this section we will analyse the boundary reconstruction error with out-vessel sensors along these three main parametric axis; sensor accuracy, proximity and density.

To start with, we perform an initial boundary reconstruction study using a 38 bidirectional probe set placed out-vessel 5cm away from the installed Mirnov coil set. The probe measurement uncertainty is swept from the typical value of a Mirnov coil (1mT) down to a hypothetical probe with a very low (1μT) measurement uncertainty.

We then study the boundary reconstruction error as function of the out-vessel sensor measurement uncertainty. The results are shown in fig. 6. We see that plasma boundary reconstruction error is reduced as the measurement uncertainty of the probe set is reduced, up to a limit (1μT) imposed by the degeneracy of the problem. Reducing the measurement uncertainty below this limit does not improve the quality of the reconstruction. The boundary reconstruction error obtained by using the standard TCV sensor configuration (38 Flux loops of 0.5mWb accuracy + 38 Mirnov coils of 1mT measurement uncertainty) is in the order of 1mm, and is also shown in fig. 6 to illustrate a typical boundary reconstruction error. Note that in the reference case the Mirnov coils are placed about 5cm closer to the plasma, and the reconstruction has also additional information from flux loops. To obtain 1mm error on plasma boundary reconstruction using only out-vessel sensors placed 5cm away from the Mirnov coil location, the out-vessel sensors will have to have a measurement uncertainty of about 0.5mT.

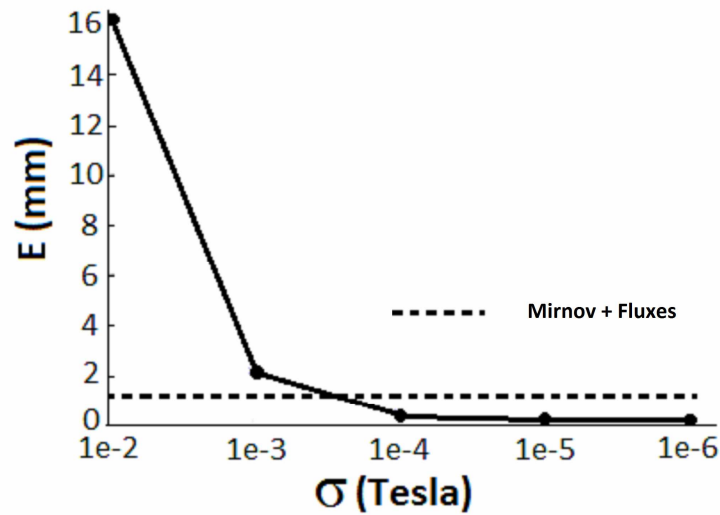

Fig. 6   Averaged boundary error reconstruction E as function of measurement uncertainty for an array of 38 sensors placed out vessel 5 cm away from the Mirnov coil location. The average is performed over all equilibria in the database for the case when there is no vessel current. The reconstruction error obtained using the standard Mirnov+ flux loop sensors is shown for comparison. The black dots correspond to the computed cases.

We can also see how the reconstruction error increases as the sensor array is positioned further away from the Mirnov coil array (fig 7) . An out-vessel sensor configuration consisting of an array 38 sensors with just 0.1mT measurement uncertainty placed at 5cm from the reference Mirnov perimeter is enough to obtain a boundary reconstruction error of 0.8 mm, which is better than the reconstruction error obtained when the standard Mirnov and flux loops sensor configuration is used (1.2mm).

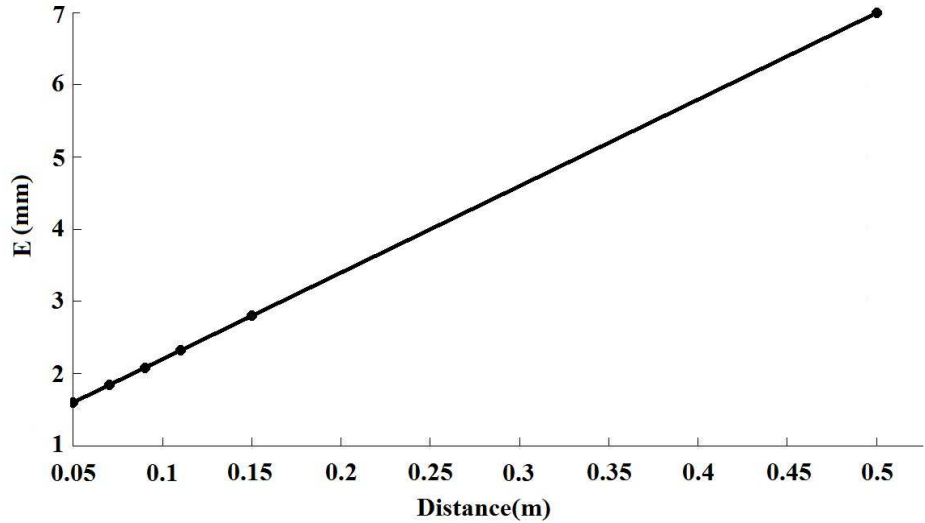

Fig. 7 Averaged boundary error reconstruction E versus the sensor array distance to the Mirnov coil array. The measurement uncertainty is 1mT and the number of sensors is 38. The average is performed over all equilibria in the database for the case when there is no vessel current. The black dots correspond to the computed cases.

We finally show in fig. 8 how the boundary reconstruction error decreases as the out-vessel sensor density is increased. The out-vessel sensor measurement uncertainty is fixed to 1mT, and the arrays are positioned at a fixed perimeter placed 5cm away from the Mirnov coils. In this case, to reduce the boundary reconstruction below the standard Mirnov/flux loop system, (1mT, 0,5mWb) an excess of about 110 out-vessel sensors would have to be used.

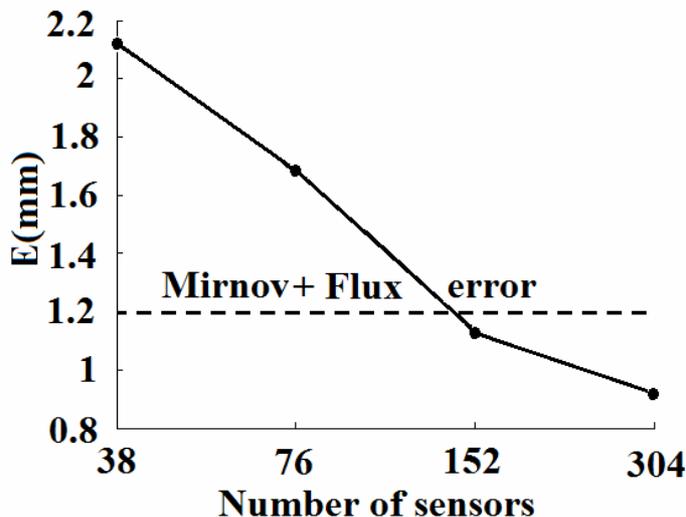

Fig. 8 Averaged boundary reconstruction error E versus number of sensors. The out vessel sensor measurement uncertainty is 1mT and the sensor distance to the Mirnov coil array is 5 cm. The reconstruction error obtained using the standard Mirnov+ flux loop sensors is shown for comparison. The average is performed over all equilibria in the database for the case when there is no vessel current. The black dots correspond to the computed cases.

To show the boundary reconstruction at work, we show in the following an analysis example for a potentially complicated boundary reconstruction from the database (inverse triangularity) without vessel currents. This is just one of the many equilibria contained in the database. Two sensor configurations are studied. The first sensor configuration corresponds to the standard set of Mirnov/ flux loop set, with accuracies of 1mT and 0.5mWb respectively, which gives a boundary reconstruction error of about 1.2 mm, as shown in fig. 6. The second sensor configuration corresponds to a hypothetically accurate (1μT measurement uncertainty) array of 38 out-vessel bidirectional sensors placed 5 cm away from the Mirnov coil reference position, without Mirnov or flux loops. The 1μT measurement uncertainty is chosen to correspond with the highest sensor accuracy used in the analysis examples of this section, as shown in fig. 6.

The boundary reconstruction results for both cases are shown in figure 9 a,b. The reference boundary is shown in blue and the reconstructed boundary is shown in red for both cases. For this particular case, it is clear that sensor accuracy and distance are traded for one another. The out-vessel sensor array can be used to reconstruct the boundary with almost 10 times better accuracy (0.1 mm average error) than the Mirnov array (1.2 mm average error), despite being placed further away from the plasma, but at the expense of using a much more accurate sensor set. The boundary reconstruction error obtained in this case is likely to be below the meaningful definition of a boundary when 3D effects are considered, and must be viewed as an ideal boundary reconstruction error. Still is a useful index to compare different sensor options as function of the array size, accuracy and positioning.

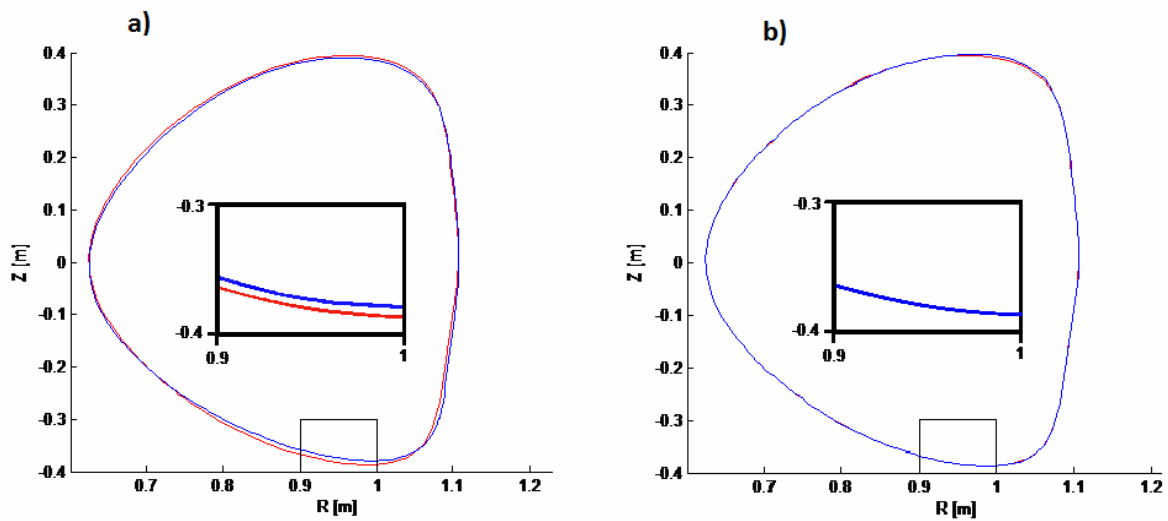

Fig. 9   Boundary reconstructions examples for a negative triangularity boundary. a) Boundary reconstruction using standard Mirnov/flux loops (1mT,0.5mWb) assuming zero vessel currents. b) The same from using only 38 outvessel sensors, without Mirnov or flux loops, 5 cm away from the original Mirnov positions with a accuracy measurement uncertainty of 10-6 Tesla. Red is true boundary, blue reconstructed. A zoomed detail is included for both cases.

As a conclusion for this section, we observe that boundary reconstruction improves with out-vessel sensor accuracy, proximity to the plasma, and density. By increasing the out-vessel sensor accuracy, proximity or density, is possible to obtain a boundary reconstruction error that is similar to that obtained the standard (Mirnov+ flux loops) sensor configuration, even when for all the studied cases the out-vessel array is placed at least 5cm further from the Mirnov coil reference position. This preliminary analysis suggests that, when there is no current induced in the vessel, sensor accuracy, density and proximity can be traded for one another. This will be confirmed in section 12, where a full parameter scan will be performed.

## 11. Analysis examples when there is current induced in the vacuum vessel

To have a reference for the boundary reconstruction error when there are currents induced in the vacuum vessel, we first start by obtaining a boundary reconstruction reference case using the standard sensor configuration of Mirnov and Flux loops, as we did in the previous section.

Fig. 10 shows the boundary reconstruction error for a set of equilibrium samples drawn randomly from the database. The average reconstruction error for the set is 4.7 mm. The larger errors that can be seen correspond  to low area and low plasma current plasmas, typical from the start and termination phases of the tokamak discharge. When the plasma area is small, the distance from sensors to plasma increases, and as we saw in the previous section for simplified cases without vessel current, this results in larger reconstruction errors. In addition, during the start and termination phases of the discharge the induced current in the vessel can be quite large, resulting in an ambiguity of solutions when resolving the spatial separation between plasma region and vessel shell region.

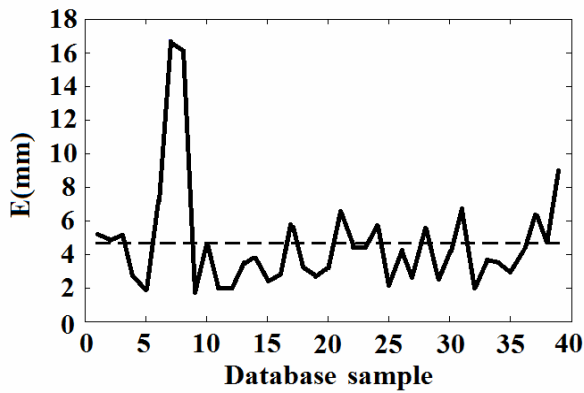

Fig. 10   Averaged boundary error reconstruction E using the standard set of Mirnov (1mT accuracy) and flux loops (0.5mWb accuracy), for several equilibria cases with vessel current. The dashed line corresponds to the average reconstruction error, which is 4.7 mm.

Another reference case of interest is the reconstruction error for the same database set using only the flux loops set, without Mirnov coils (fig. 11). The flux loops in this case are a type of out-vessel sensor (for which we have measurements available in TCV) that is useful to compare with other, more accurate, virtual out-vessel sensor options that we have to simulate. The average error in this case is 10mm, and as in the previous case the larger reconstruction errors that can observed in the graph correspond to cases of low plasma current and cross-section.

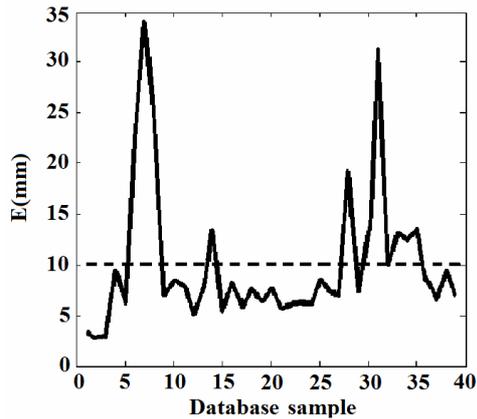

Fig. 11 Averaged boundary error reconstruction E using only the standard set of flux loops, for several equilibria cases with vessel current. The dashed line corresponds to the average reconstruction error, which is 10 mm.

Next, we determine the average boundary reconstruction error for a sensor configuration consisting of 38 out-vessel sensors placed 5cm away from the reference Mirnov coil positions. The measurement uncertainty of the out-vessel sensors has been changed between $10^{-1}$ and $10^{-6}$ Tesla (Fig 12). There is a very little change, indicating that the information about the field at those positions and sensor directions are not enough to disambiguate the solution space much further. The general trend is a very weak decrease in the reconstruction error when the measurement uncertainty is decreased. The average reconstruction error obtained when using out-vessel sensors (9.7mm) is equivalent to that obtained by using flux loops alone (10mm). This error is two times larger than the boundary reconstruction error obtained by using the standard set of in-vessel Mirnov and out-vessel fluxes, which was 4.7mm. It is clear that the information provided by the Mirnov coils helps to disambiguate the solution space, since in this case the vessel current is sandwiched between two sets of sensors, Mirnov coils and fluxes.

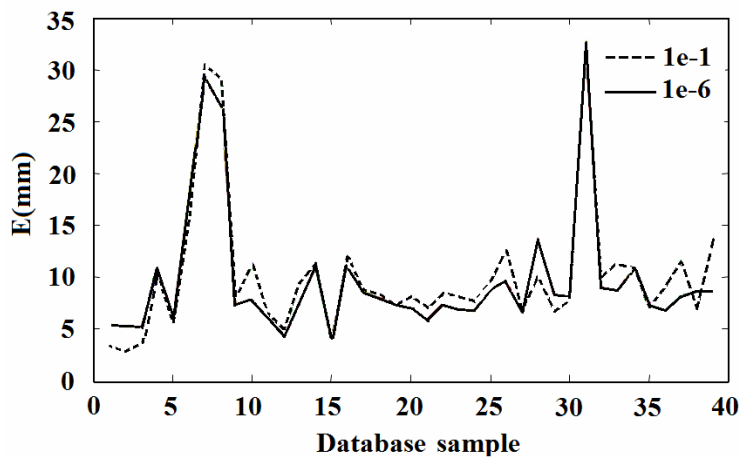

Fig. 12 Averaged boundary error reconstruction E using only 38 out vessel sensors and a subset of equilibria cases with vessel current drawn randomly from the database. The measurement uncertainty at the out vessel sensor positions (d=0.05 m) is changed between $10^{-1}$ and $10^{-6}$ Tesla. The average reconstruction error is 9.7 mm .There is no impact on the result, so it is not possible to reduce the reconstruction error below 1cm, regardless of the out-vessel sensor accuracy.

We finally check the boundary reconstruction error if we use the in-vessel Mirnov sensor set and an out-vessel sensor set (with 1mT sensor uncertainty) placed outside the vessel at d=0.05 m . The results are shown in fig 13. It shows a major improvement in reconstruction accuracy both in comparison to using only out-vessel sensors (fig 12) or flux loops (fig 11). Compared with the case when the set of Mirnov coils and flux loops (fig 10) is used, the reconstruction error is reduced by a factor of two, which comes as no surprise since the accuracy of the out-vessel sensors used in this last case exceeds that of the flux loops. In general, the best reconstructions results are obtained when in-vessel Mirnov and out-vessel sensors are used, so the vessel currents are then fully included between two boundaries with measurements, and so the vessel currents are then far easier to separate from the plasma currents.

We can also see how the reconstruction error increases as the out-vessel array is positioned further and further away from the Mirnov coil array (Fig 14 ). The out-vessel array is placed at distances between 5 cm and 20 cm from the position of the Mirnov coils. The sensor configuration used is the previous one, including both an array of out-vessel sensors outside the vacuum vessel, and a standard array of Mirnov coils inside the vacuum vessel.

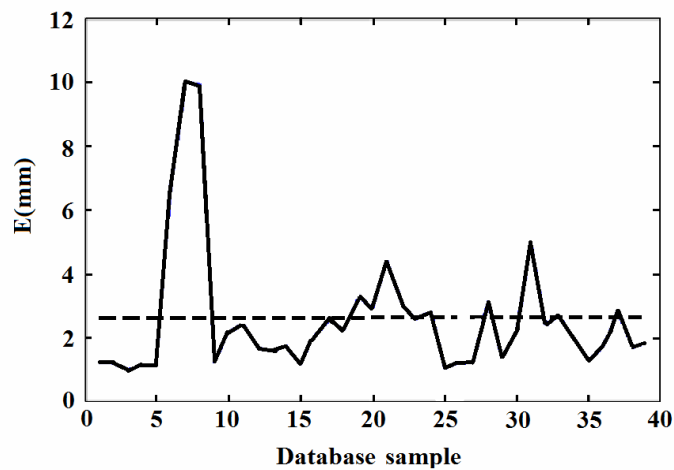

Fig. 13   Averaged boundary error reconstruction E using the whole set of Mirnov and an out-vessel sensor set on a subset of equilibria cases with vessel current drawn randomly from the database. The dashed line corresponds to the average value, 2.6mm.

There are 38 sensors in each array and the directionality of the out-vessel sensors is the same as for the Mirnov coils.   Fig.14 shows that the reconstruction error approaches the reconstruction error for the sensor configuration with only Mirnov and flux loops (4.7mm) at about 20 cm from the position of the Mirnov coils. Again, no consideration has been taken to whether there is space for the sensors in those positions; this is done to indicate the distance at which such sensors could have been placed in principle.

Practically at TCV, only distances up to about 7 cm would seem feasible. Putting the sensors too near to the vacuum vessel might make them too sensitive to details in the actual model used for the vessel currents. Furthermore, and again from a practical point of view, there is no space at TCV between the vacuum vessel and the PF coils in the high field side. In any case, the results show that there is only a small region where placing sensors would improve the boundary reconstruction substantially. Improving the accuracy of the sensors will not reduce the reconstruction error significantly if placed too far away.

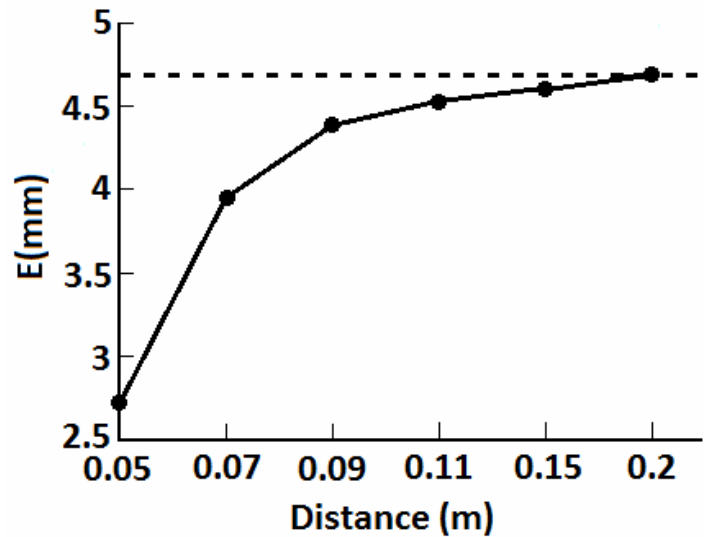

Fig. 14 Averaged boundary error reconstruction E using the whole set of Mirnov and out-vessel sensors (solid line). Far away from the vessel, the error approaches the error from the reconstruction using Mirnov and flux loops (discontinuous line). The black dots correspond to the computed cases.

## 12. Parameter scans

In this section the out-vessel sensor positioning, number of sensors, and model predictive accuracy will be scanned and the boundary reconstruction accuracy will be calculated for each parameter combination. The ranges of the parameters are shown in Table 1.

**Table** 1

| Parameter | Values |
|---|---|
| Position of out-vessel sensor array relative to the Mirnov reference | 0.05, 0.1, 0.2, 0.3, 0.4, 0.5, 0.6 [m] |
| Measurement uncertainty ($\sigma$) | $10^{-2}$, $10^{-3}$, $10^{-4}$, $10^{-5}$, $10^{-6}$ [Tesla] |
| Number of sensors | 38, 76, 152, 304 |

From the analysis carried out in the previous sensors, it is clear that installation of accurate out-vessel sensors could improve the boundary reconstruction substantially, if used in conjunction with existing in-vessel sensors and a full passive vessel model. Also, it could fully replace the standard Mirnov/flux loop combination for the standard case when vacuum vessel induced currents can be ignored, as is the case of steady state reactor operation during the flat top of the discharge. For the latter case the boundary reconstruction error can be orders of magnitude lower than the standard Mirnov/flux loop combination, depending on how well the forward model can predict the actual measurement. What the achievable boundary reconstruction accuracy would be for each case will depend on both the position of the sensors, the number of sensors, and the achievable model prediction accuracy. The latter is determined by how well the model of plasma current and poloidal field coils can predict the actual vacuum field at the sensor positions. Only an actual implementation of out-vessel sensors can determine this accuracy since, as opposed to Mirnov coils and flux loops, the measurement error with the out-vessel sensors will not be a limiting factor when comparing model prediction with measurement. In the graphs that follow, both model prediction accuracy and boundary reconstruction accuracy will be expressed logarithmically, as $-\log_{10}\sigma$, or $-\log_{10}E$, where $\sigma$ is the measurement uncertainty with which the model can predict the out-vessel sensor measurement (in Tesla), and $E$ is the average boundary reconstruction error over the database ( in metres).

Expressed in this way, larger values in the y axis correspond with more accurate sensors, while larger values in the contour lines correspond to more accurate boundary reconstructions.

The first scan (fig. 15 ) shows the results for a sensor configuration with only out-vessel sensors, with no other diagnostics used, and no model for induced vessel currents. This is suitable for the steady state, flat-top phase of the tokamak discharge, when the induced currents in the vacuum vessel can be ignored. The contour plots show the boundary reconstruction accuracy on the database for different number of sensors, as a function of model prediction accuracy and position of the out-vessel sensor array. From these plots it is clear that distance and accuracy are interchangeable in the sense that putting the sensors further away can give the same accuracy as putting the sensors near to the vacuum vessel if at the same time the accuracy is increased. This can be seen by following the different contour lines from left to right in the plots. For 76 sensors and upwards, the most optimistic case (orange coloured area with value 4) gives an error below $10^{-4}$ m=0.1mm, which is very likely far below what is physically meaningful since the boundary itself will not be so well defined when plasma 3D effects are considered. In that sense, this out-vessel sensor configuration could reach a boundary reconstruction accuracy that is at the limit of what is physically meaningful for plasma boundary inference.

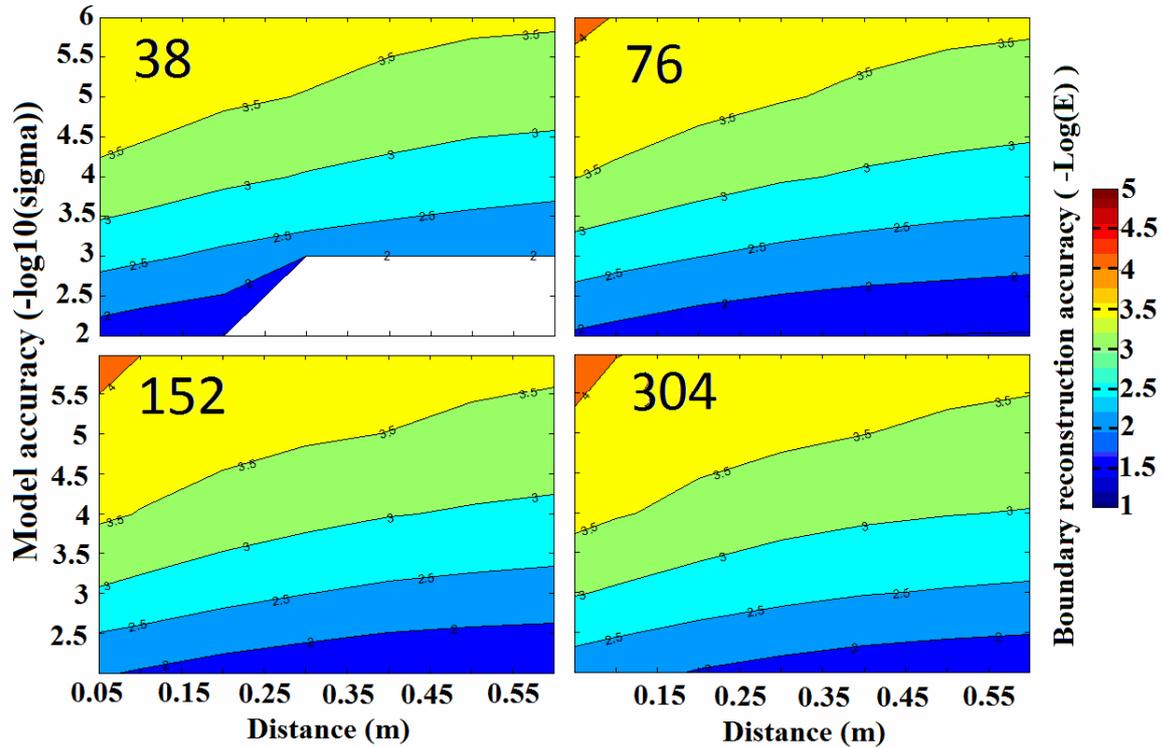

Fig. 15    Contour plots of the reconstruction accuracy as function of sensor configuration (number of sensors, accuracy and distance to the reference Mirnov coil set). Sensor and boundary reconstruction accuracies are in logarithmic scale. A sensor configuration consisting of out-vessel sensors is used. No vessel currents model is considered. Larger values in the y axis correspond with more accurate sensors, while larger values in the contour lines correspond to more accurate boundary reconstructions.

Fig. 16 shows the case of boundary reconstruction using a sensor configuration consisting of out-vessel sensors and a model for the vessel currents. This is suitable for the plasma current ramp-up and ramp-down transients of the discharge, when the induced currents in the vacuum vessel are expected to be relevant. The achievable accuracy for this case is as expected much lower, as we had already seen in the examples of the previous section. The boundary reconstruction error in the dark coloured area with accuracy value 2 is $10^{-2}$m=10mm. At the closest distance to the plasma, d=0.05 in this study, this requires a model accuracy of 3. Recalling that we are using logarithmic scale, this corresponds with $10^{-3}$T=1mT. In this case, the boundary reconstruction accuracy doesn't improve when model accuracy is increased. This is an indication that the main source of uncertainty is the ambiguity of solutions related to the necessity of doing simultaneous inference on two spatial regions: the vessel shell and the plasma region. There is a slight increase in reconstruction accuracy when the number of sensors are increased, but the reconstruction error stays at about $10^{-2}$, or 10mm. This is the same reconstruction error obtained by using flux loops alone, as it was found in section 11.

So we wonder if a sensor configuration with Mirnov coils and out-vessel sensors could give any advantage with respect to using the standard sensor configuration of Mirnov coils and flux loops alone. This is answered in the next parameter scan. We show in fig.17 the improvement in boundary reconstruction accuracy that is possible when a hybrid sensor configuration consisting of out-vessel sensors, Mirnov coils and flux loops is used. Vessel currents are inferred using our vessel current model, so this sensor configuration and model is in principle suitable for the steady state and transient phases of the discharge.

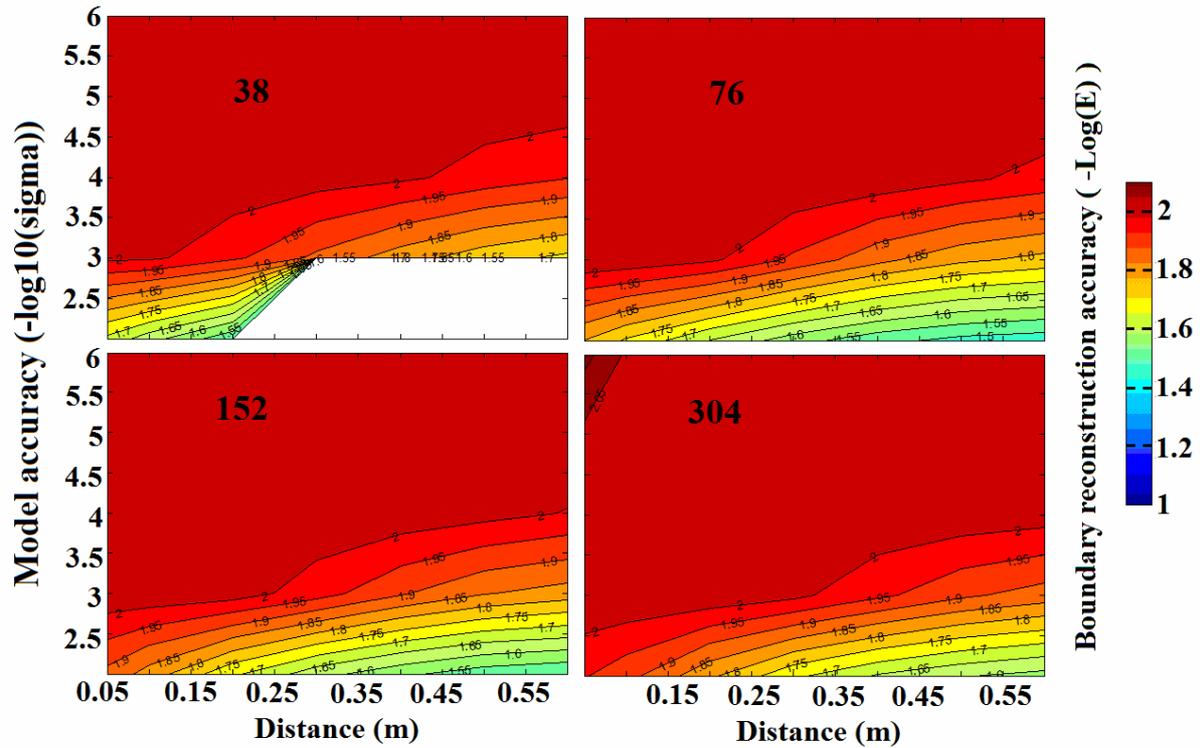

Fig. 16   Contour plots of the reconstruction accuracy as function of sensor configuration (number of sensors, accuracy and distance to the reference Mirnov coil set). Sensor and boundary reconstruction accuracies are in logarithmic scale. A sensor configuration consisting of out-vessel sensors and a model with vessel current are used. Larger values in the y axis correspond with more accurate sensors, while larger values in the contour lines correspond to more accurate boundary reconstructions.

In the configuration with 38 sensors, the boundary reconstruction accuracy represented by the cyan coloured contour (value 2.6) corresponds with a boundary reconstruction error of $10^{-2.6}$m = 2.5mm. This reconstruction error is about two times smaller than the reconstruction error achievable with the sensor configuration of Mirnov coils and flux loops (Fig. 10) , which was previously estimated in 4.7mm.

The intersection of this contour line with the vertical axis at 3.5 corresponds with sensors positioned 5 cm from the original Mirnov coils (distance=0.05) and a model accuracy of

$10^{-3.5}$T=0.3mT. This measurement uncertainty can be achieved with standard Hall sensors.

The achievable boundary reconstruction accuracy can be further improved by increasing the number of sensors. For instance, using 304 sensors distanced 5cm from the Mirnov coils with 0.3mT measurement uncertainty, we can get a 0.8mm boundary reconstruction error, which is probably at the limit of what is physically meaningful, since the boundary itself may not be so well defined when plasma 3D effects are considered.

We could also choose to keep a boundary reconstruction error of 2.5 mm (cyan contour) and try at the same time to place the sensors as far away as possible. We could place, for instance, 304 sensors at 17 cm from the Mirnov reference, but we would need then an extremely low measurement uncertainty of $10^{-6}$T.

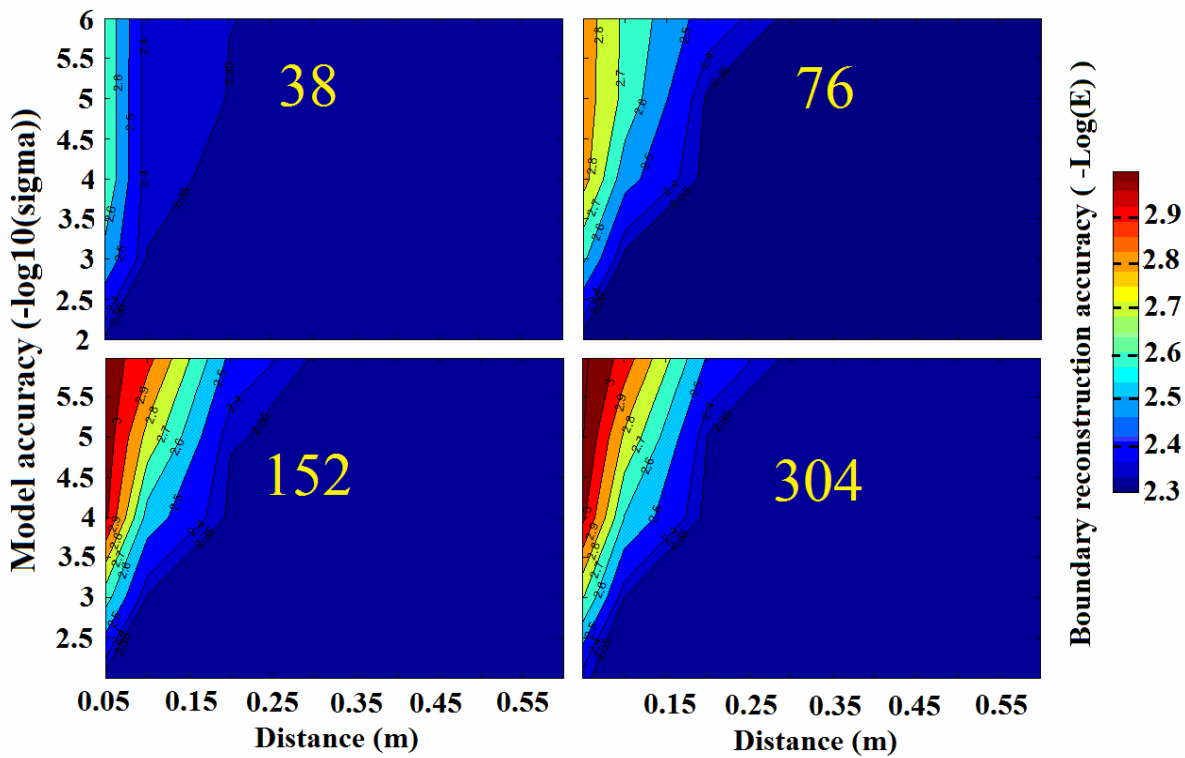

Fig. 17  Contour plots of the reconstruction accuracy as function of sensor configuration (number of sensors, accuracy and distance to the reference Mirnov coil set). Sensor and boundary reconstruction accuracies are in logarithmic scale. A sensor configuration consisting of out-vessel sensors, Mirnov coils and flux loops is used in combination with a model that includes vessel currents. Larger values in the y axis correspond with more accurate sensors, while larger values in the contour lines correspond to more accurate boundary reconstructions.

## 13. Applicability to a tokamak fusion reactor

The study presented has been developed using a small tokamak as test bed. However, the issues related to equilibrium reconstruction are not fundamentally different from any other air core tokamak reactor regardless of its size, as long as the machine aspect ratio and plasma shapes are comparable. In this respect the TCV tokamak plasmas are very similar to the expected plasma shapes of future tokamak reactors. One important difference between TCV and a future reactor, though, will likely be the vacuum vessel itself. TCV has a large vacuum vessel compared with the plasma size, to be able to accommodate and study a large variety of plasma shapes and divertor configurations. This means that the magnetic sensors have to be located quite far from the plasma, particularly at the top or bottom regions of the vacuum vessel. This quite likely will not be the case in a nuclear fusion tokamak reactor, where increasing the vessel size will have an impact on magnet size and machine cost.

Nevertheless, if we assume the results of this study can be extrapolated to a fusion reactor, a hybrid configuration with Mirnov and out-vessel sensors would be likely to work for the following qualitative reasons:

1) The Mirnov coil array would not be influenced by radiation during the ramp–up phase, since reactor conditions have not been reached yet in this phase. The combination of Mirnov and out vessel sensors could give in this phase very accurate plasma boundary reconstructions.

2) The out-vessel sensors would be enough to reconstruct the equilibrium during the flat top phase; the unreliable Mirnov measurements during this radiation active phase are not required for the equilibrium reconstruction and can be ignored if necessary.

However, to have a quantitative measure of the required number of sensors, accuracy and proximity to the plasma for a future tokamak reactor such as ITER (for which experimental data does not exist) it would be necessary to repeat these parameter scans for the new machine geometry using synthetic tokamak plasma equilibria, and select a suitable configuration in the parameter scan diagrams starting from a specification of what a safe distance would be in terms of the admissible radiation levels at the out-vessel sensor location.

## 14. Conclusions

We have studied the tokamak equilibrium problem with out-vessel magnetic sensors for transient and steady state phases of the tokamak, as function of sensor density, accuracy and distance to the vacuum vessel.

We find that in the standard case when the current induced in the vacuum vessel is negligible (e.g. flat-top of the discharge), the increase of sensor proximity to the vacuum vessel, and/or sensor accuracy, and/or sensor density in the array increases the quality of the reconstructed boundary. Sensor accuracy, density and proximity to the vessel can be traded for one another to obtain a desired level of accuracy in the reconstructed boundary. As a practical example, we have shown that, during the flat-top phase, out-vessel sensors with moderate accuracies the order of 0.3mT can fully replace the standard sensor configuration of Mirnov and flux loops, obtaining a similar accuracy on the reconstructed boundary.

However, when the current induced in the vacuum vessel is important (*transient phases*), out-vessel sensors alone are incapable to reconstruct the plasma boundary with an error of less than 1cm. This holds equally for flux loops or any type of out vessel sensor, being almost independent of sensor proximity, density or accuracy. This is an indication that the main source of uncertainty in this case is the ambiguity of solutions corresponding to vessel and plasma regions, which can not be fully discriminated using out-vessel measurements alone.

To obtain a good equilibrium reconstruction when there are currents induced in the vessel, it is necessary to have two sensor arrays, one inside and one outside the vessel, just like the combination of Mirnov coils and flux loops. This helps to disambiguate the space of solutions corresponding to vessel and plasma current distribution regions, as the vessel current is then sandwiched between two layers of measurements. Sensor configurations consisting of Mirnov coils and out-vessel sensors give the best boundary reconstructions in this case.

We find that, in general, there is a broad region in the parameter space where different combinations of sensor accuracy, density and proximity to the plasma give a similar accuracy in the reconstructed boundary, up to some limit. In case of TCV, the furthest away a sensor array can be placed to obtain a boundary reconstruction error of 2.5 mm is about 17cm, using 304 sensors with a measurement uncertainty of 10-6 Tesla. Increasing sensor density and/or reducing measurement uncertainty beyond these values does not reduce the boundary reconstruction error further.

Extrapolation of the results to a tokamak reactor suggests that a hybrid configuration with sensors inside and outside the vacuum vessel could be used to obtain a good boundary reconstruction during both the transient and the flat-top of the discharges, if out vessel magnetic sensors of sufficient density and accuracy can be placed sufficiently far outside the vessel to minimize radiation damage.

## Acknowledgements

This work has been supported by the Euratom mobility programme, the EU FP7 EFDA under the task WP09-DIA-02-01 WP III-2-c, the University of the Basque Country (UPV/EHU) through Research Project GIU11/02, and the Spanish Ministry of Science and Innovation (MICINN) through Research Project ENE2010-18345. This work was possible thanks to a short stay at CRRP in which access to the TCV database was given.

# Appendix : TCV coil and magnetic sensor parameters

**Table** 2. PF coil beam positions, dimensions, number of turns, and circuit.

| Coil name | Centre      R [m] | Centre      Z [m] | Width [m] | Height [m] | Turns | Circuit |
|-----------|-------------------|-------------------|-----------|------------|-------|---------|
| A_001 | 0.4225 | 0 | 0.063 | 1.584 | 143 | OH_001 |
| B_001 | 0.4457 | -0.936 | 0.1095 | 0.188 | 29 | OH_002 |
| B_002 | 0.4457 | 0.936 | 0.1095 | 0.188 | 29 | OH_002 |
| C_001 | 0.6215 | -1.11 | 0.066 | 0.12 | 12 | OH_002 |
| C_002 | 0.6215 | 1.11 | 0.066 | 0.12 | 12 | OH_002 |
| D_001 | 1.1765 | -1.17 | 0.066 | 0.09 | 8 | OH_002 |
| D_002 | 1.1765 | 1.17 | 0.066 | 0.09 | 8 | OH_002 |
| E_001 | 0.5050 | -0.7 | 0.051 | 0.18 | 34 | E_001 |
| E_002 | 0.5050 | -0.5 | 0.051 | 0.18 | 34 | E_002 |
| E_003 | 0.5050 | -0.3 | 0.051 | 0.18 | 34 | E_003 |
| E_004 | 0.5050 | -0.1 | 0.051 | 0.18 | 34 | E_004 |
| E_005 | 0.5050 | 0.1 | 0.051 | 0.18 | 34 | E_005 |
| E_006 | 0.5050 | 0.3 | 0.051 | 0.18 | 34 | E_006 |
| E_007 | 0.5050 | 0.5 | 0.051 | 0.18 | 34 | E_007 |
| E_008 | 0.5050 | 0.7 | 0.051 | 0.18 | 34 | E_008 |
| F_001 | 1.3095 | -0.77 | 0.1197 | 0.0748 | 36 | F_001 |
| F_002 | 1.3095 | -0.61 | 0.1197 | 0.0748 | 36 | F_002 |
| F_003 | 1.3095 | -0.31 | 0.1197 | 0.0748 | 36 | F_003 |
| F_004 | 1.3095 | -0.15 | 0.1197 | 0.0748 | 36 | F_004 |
| F_005 | 1.3095 | 0.15 | 0.1197 | 0.0748 | 36 | F_005 |
| F_006 | 1.3095 | 0.31 | 0.1197 | 0.0748 | 36 | F_006 |
| F_007 | 1.3095 | 0.61 | 0.1197 | 0.0748 | 36 | F_007 |
| F_008 | 1.3095 | 0.77 | 0.1197 | 0.0748 | 36 | F_008 |
| G_001 | 1.0986 | -0.651 | 0.005 | 0.005 | 1 | G_001 |
| G_002 | 1.1140 | -0.633 | 0.005 | 0.005 | 1 | G_001 |
| G_003 | 1.1294 | -0.615 | 0.005 | 0.005 | 1 | G_001 |
| G_004 | 1.1294 | 0.615 | 0.005 | 0.005 | 1 | G_001 |
| G_005 | 1.1140 | 0.633 | 0.005 | 0.005 | 1 | G_001 |
| G_006 | 1.0985 | 0.651 | 0.005 | 0.005 | 1 | G_001 |
| T_001 | 1.5540 | -0.78 | 0.03 | 0.15 | 1 | T_001 |
| T_002 | 1.7170 | -0.78 | 0.03 | 0.15 | 1 | T_001 |
| T_003 | 1.7540 | -0.78 | 0.03 | 0.15 | 1 | T_001 |

**Table** 3. PF circuits formed by linear combinations of coil currents. After each coil is indicated in parenthesis the amount and sign of the circuit current that goes through the given coil.

| Circuit name | Coils(weight) |
|---|---|
| OH_001 | A_001(1) |
| OH_002 | B_001(1), B_002(1), C_001(1), C_002(1), D_001(1), D_002(1) |
| E_001 | E_001(1) |
| E_002 | E_002(1) |
| E_003 | E_003(1) |
| E_004 | E_004(1) |
| E_005 | E_005(1) |
| E_006 | E_006(1) |
| E_007 | E_007(1) |
| E_008 | E_008(1) |
| F_001 | F_001(1) |
| F_002 | F_002(1) |
| F_003 | F_003(1) |
| F_004 | F_004(1) |
| F_005 | F_005(1) |
| F_006 | F_006(1) |
| F_007 | F_007(1) |
| F_008 | F_008(1) |
| G_001 | G_001(1), G_002(1), G_003(1), G_004(-1), G_005(-1), G_006(-1) |
| T_001 | T_001(-26/68), T_002(-42/68), T_003(1) |

**Table** 4. Mirnov coil positions and poloidal directions.

| Coil | R [m] | Z [m] | Angle [radians] |
|---|---|---|---|
| 1 | 0.606 | 0.000 | 1.571 |
| 2 | 0.606 | 0.115 | 1.571 |
| 3 | 0.606 | 0.230 | 1.571 |
| 4 | 0.606 | 0.345 | 1.571 |
| 5 | 0.606 | 0.460 | 1.571 |
| 6 | 0.606 | 0.575 | 1.571 |
| 7 | 0.606 | 0.690 | 1.571 |
| 8 | 0.629 | 0.746 | 0.785 |
| 9 | 0.695 | 0.764 | 0.000 |
| 10 | 0.790 | 0.764 | 0.000 |
| 11 | 0.885 | 0.764 | 0.000 |
| 12 | 0.980 | 0.764 | 0.000 |
| 13 | 1.042 | 0.735 | -0.698 |
| 14 | 1.119 | 0.671 | -0.698 |
| 15 | 1.154 | 0.575 | -1.571 |
| 16 | 1.154 | 0.460 | -1.571 |
| 17 | 1.154 | 0.345 | -1.571 |

| 18 | 1.154 | 0.230 | -1.571 |
| 19 | 1.154 | 0.115 | -1.571 |
| 20 | 1.154 | 0.000 | -1.571 |
| 21 | 1.154 | -0.115 | -1.571 |
| 22 | 1.154 | -0.230 | -1.571 |
| 23 | 1.154 | -0.345 | -1.571 |
| 24 | 1.154 | -0.460 | -1.571 |
| 25 | 1.154 | -0.575 | -1.571 |
| 26 | 1.119 | -0.671 | -2.443 |
| 27 | 1.042 | -0.735 | -2.443 |
| 28 | 0.980 | -0.764 | -3.142 |
| 29 | 0.885 | -0.764 | -3.142 |
| 30 | 0.790 | -0.764 | -3.142 |
| 31 | 0.695 | -0.764 | -3.142 |
| 32 | 0.629 | -0.746 | 2.356 |
| 33 | 0.606 | -0.690 | 1.571 |
| 34 | 0.606 | -0.575 | 1.571 |
| 35 | 0.606 | -0.460 | 1.571 |
| 36 | 0.606 | -0.345 | 1.571 |
| 37 | 0.606 | -0.230 | 1.571 |
| 38 | 0.606 | -0.115 | 1.571 |

**Table** 5. Ideal flux loop (1-38) positions

| Coil | R [m] | Z [m] |
| --- | --- | --- |
| 1 | 0.584 | 0.000 |
| 2 | 0.584 | 0.115 |
| 3 | 0.584 | 0.230 |
| 4 | 0.584 | 0.345 |
| 5 | 0.584 | 0.460 |
| 6 | 0.584 | 0.575 |
| 7 | 0.584 | 0.690 |
| 8 | 0.600 | 0.772 |
| 9 | 0.695 | 0.791 |
| 10 | 0.790 | 0.791 |
| 11 | 0.885 | 0.791 |
| 12 | 0.980 | 0.791 |
| 13 | 1.059 | 0.756 |
| 14 | 1.136 | 0.691 |
| 15 | 1.176 | 0.575 |
| 16 | 1.176 | 0.460 |
| 17 | 1.176 | 0.345 |
| 18 | 1.176 | 0.230 |
| 19 | 1.176 | 0.115 |

| 20 | 1.176 | 0.000 |
|----|-------|-------|
| 21 | 1.176 | -0.115 |
| 22 | 1.176 | -0.230 |
| 23 | 1.176 | -0.345 |
| 24 | 1.176 | -0.460 |
| 25 | 1.176 | -0.575 |
| 26 | 1.136 | -0.691 |
| 27 | 1.059 | -0.756 |
| 28 | 0.980 | -0.791 |
| 29 | 0.885 | -0.791 |
| 30 | 0.790 | -0.791 |
| 31 | 0.695 | -0.791 |
| 32 | 0.600 | -0.772 |
| 33 | 0.584 | -0.690 |
| 34 | 0.584 | -0.575 |
| 35 | 0.584 | -0.460 |
| 36 | 0.584 | -0.345 |
| 37 | 0.584 | -0.230 |
| 38 | 0.584 | -0.115 |

**Table** 6. Coil flux loop (39-61) positions

| 39 | 0.446 | 1.076 |
|----|-------|-------|
| 40 | 0.500 | -0.936 |
| 41 | 0.500 | 0.936 |
| 42 | 0.655 | -1.110 |
| 43 | 0.655 | 1.110 |
| 44 | 1.143 | -1.170 |
| 45 | 1.143 | 1.170 |
| 46 | 0.530 | -0.700 |
| 47 | 0.530 | -0.500 |
| 48 | 0.530 | -0.300 |
| 49 | 0.530 | -0.100 |
| 50 | 0.530 | 0.100 |
| 51 | 0.530 | 0.300 |
| 52 | 0.530 | 0.500 |
| 53 | 0.530 | 0.700 |
| 54 | 1.250 | -0.770 |
| 55 | 1.250 | -0.610 |
| 56 | 1.250 | -0.310 |
| 57 | 1.250 | -0.150 |
| 58 | 1.250 | 0.150 |
| 59 | 1.250 | 0.310 |
| 60 | 1.250 | 0.610 |
| 61 | 1.250 | 0.770 |

**Table** 7. Pulses in the full database together with time limits within those pulses within which equilibria has been randomly selected.

| Pulse | t0 [s] | t1 [s] | Samples |
|-------|--------|--------|---------|
| 11368 | 0.040 | 1.320 | 30 |
| 11928 | 0.050 | 0.670 | 30 |
| 36151 | 0.050 | 0.634 | 23 |
| 39950 | 0.021 | 1.543 | 30 |
| 39952 | 0.021 | 1.549 | 30 |
| 41450 | 0.100 | 2.160 | 30 |
| 41452 | 0.100 | 2.155 | 30 |
| 41453 | 0.100 | 2.305 | 30 |
| 41456 | 0.110 | 1.010 | 30 |
| 41479 | 0.050 | 2.460 | 30 |
| 41481 | 0.050 | 1.760 | 30 |
| 41834 | 0.050 | 2.210 | 30 |
| 42345 | 0.050 | 2.210 | 30 |